\documentclass{appolb}
\usepackage{graphicx}
\usepackage[T1]{fontenc}

\usepackage{amsmath,amssymb}

\usepackage[font=small,skip=0pt]{caption}

\graphicspath{{figures/}}

\usepackage{multicol}

\usepackage{xcolor}
\usepackage{tikz}
\usetikzlibrary{decorations.pathreplacing,calc}
\usetikzlibrary{shapes.misc, positioning}
\usetikzlibrary{fadings}
\usetikzlibrary{backgrounds}

\usetikzlibrary{arrows,calc}
\usetikzlibrary{shapes.misc, positioning}
\usetikzlibrary{positioning,decorations.text}
\usetikzlibrary{shapes.geometric}
\usetikzlibrary{shapes.symbols}

\definecolor{olive}{rgb}{0.3, 0.4, .1}
\definecolor{fore}{RGB}{249,242,215}
\definecolor{back}{RGB}{51,51,51}
\definecolor{title}{RGB}{255,0,90}
\definecolor{dgreen}{rgb}{0.,0.6,0.}
\definecolor{gold}{rgb}{1.,0.84,0.}
\definecolor{JungleGreen}{cmyk}{0.99,0,0.52,0}
\definecolor{BlueGreen}{cmyk}{0.85,0,0.33,0}
\definecolor{RawSienna}{cmyk}{0,0.72,1,0.45}
\definecolor{Magenta}{cmyk}{0,1,0,0}

\definecolor{bblue}{rgb}{0,0.3,0.45}
\definecolor{dblue}{rgb}{0,0.2,0.3}
\definecolor{bblue1}{rgb}{0.0,0.45,0.7}
\definecolor{bblue2}{rgb}{0.01,0.37,0.52}
\definecolor{bblue3}{rgb}{0.72,0.81,0.85}
\definecolor{ggreen}{rgb}{0,0.292372549002,0.0}
\definecolor{rred}{cmyk}{0, 0.919, 0.856, 0.702}
\definecolor{ggray}{rgb}{0.35,0.35,0.25}
\definecolor{mgray}{rgb}{0.55,0.55,0.45}
\definecolor{lgray}{rgb}{0.85,0.90,0.90}

\definecolor{lblue}{RGB}{180,200,210}
\definecolor{llblue}{RGB}{227,235,242}
\definecolor{lllblue}{RGB}{241,245,248}
\definecolor{mblue}{RGB}{46,108,139}
\definecolor{rrred}{RGB}{170,20,10}

\definecolor{cblue}{HTML}{004D73}
\definecolor{cred}{HTML}{931621}
\definecolor{cyellow}{HTML}{F4D35E}
\definecolor{cgreen}{HTML}{AFC97E}
\definecolor{cpurple}{HTML}{493548}
\definecolor{cmblue}{HTML}{8BAEBF}
\definecolor{cmred}{HTML}{FF958A}
\definecolor{cmyellow}{HTML}{FAEBB5}
\definecolor{cmgreen}{HTML}{DAE6C4}
\definecolor{cmpurple}{HTML}{ACA3AB}
\definecolor{clblue}{HTML}{B9CED8}
\definecolor{clred}{HTML}{FFC1B2}
\definecolor{clyellow}{HTML}{FCF3D3}
\definecolor{clgreen}{HTML}{E9F0DB}
\definecolor{clpurple}{HTML}{CDC7CD}
\definecolor{cdblue}{HTML}{003D53}
\definecolor{cdred}{HTML}{831215}
\definecolor{cdyellow}{HTML}{945315}
\definecolor{cdgreen}{HTML}{5F873E}
\definecolor{cdpurple}{HTML}{392538}

\newcommand{\bi}{\begin{itemize}}
	\newcommand{\ei}{\end{itemize}}
\newcommand{\ig}{\includegraphics}

\usepackage{xspace}
\newcommand{\na}{\textsc{Na61/Shine}\xspace}
\newcommand{\au}{\textcolor{rrred}{Au+Au}\xspace}
\newcommand{\pb}{\textcolor{rrred}{Pb+Pb}\xspace}
\newcommand{\ar}{\textcolor{orange!80!rred}{Ar+Sc}\xspace}
\newcommand{\be}{\textcolor{ggreen}{Be+Be}\xspace}
\newcommand{\pp}{\textcolor{bblue}{p+p}\xspace}

\begin{document}
\title{On strangeness in NA61/SHINE
\thanks{Presented at  Excited QCD, Kopaonik, Serbia,  11-15 March 2018}%
}

\author{Maciej P. Lewicki\\ {\small for the NA61/SHINE Collaboration}
\address{Institute of Theoretical Physics, Univeristy of Wroclaw}
}
\maketitle
\begin{abstract}
NA61/SHINE is a fixed target experiment at the CERN Super-Proton-Synchrotron. The main goals of the experiment are to discover the critical point of strongly interacting matter and to study the properties of the onset of deconfinement. In order to reach these goals, a study of hadron production properties is performed in nucleus-nucleus, proton-proton and proton-nucleus interactions as a function of collision energy and size of the colliding nuclei. 
In this paper, I will review recent results on strangeness production in p+p, Be+Be and Ar+Sc collisions in the SPS energy range. Kinematic spectra and mean multiplicities of kaons obtained with various analysis methods will be shown. An overview of strangeness production and its dependence on system size in the vicinity of the phase transition will be presented as well.
\PACS{25.75.-q, 25.75.Ag, 25.75.Gz, 25.75.Dw, 25.75.Nq}
\end{abstract}

\section{Study of strong interactions in NA61/SHINE}
NA61/SHINE is a fixed target spectrometer \cite{facility} located in CERN's North Area, utilizing the SPS proton, ion and hadron beams. Tracking capabilities are provided by four large volume Time Projection Chambers (TPC), two of which are located in magnetic fields. The Projectile Spectator Detector (PSD), a zero degree, modular calorimeter, is used to determine the centrality of the collisions.

The aim of the experiment is to explore the QCD phase diagram $(\mu_B, T)$ by a two-dimensional scan in collision energy and system size. The yields of hadrons produced in the collisions are studied for indications of the onset of deconfinement and the critical point of the phase transition. This paper discusses the properties of strangeness production in the vicinity of the deconfinement phase transition. Special attention is given to recent results on Be+Be and Ar+Sc collisions, which provide new insight into the system size dependence of hadron production near the onset of deconfinement.

\section{Strangeness production near deconfinement phase transition}
There is no net strangeness content in the colliding nuclei, so any (anti-)strange particles recorded in the detectors are produced by strong interactions during the collision. Therefore (anti-)strange particles serve as an excellent probe of the phase transition dynamics.

In the hadronic phase the lightest strangeness carriers are $K$ mesons, the mass ($M_K\approx500$~MeV) of which largely exceeds the critical temperature of hadron matter, estimated at around 160~MeV. In the QGP phase the role of lightest strangeness carriers is ceded to relatively light strange and anti-strange quarks ($m_s\approx$100MeV).

In the statistical approach (as in \cite{smes}) we may predict the dependence of strangeness production on collision energy by invoking the Boltzmann approximation as an intuitive toy model:
$$
\langle N_i \rangle = \frac{gV}{(2\pi)^3}\int d^3p\frac{1}{e^{E/T}\pm1}
$$
leading to the following yield estimates for heavy and light particles:
$$
\textrm{\small(heavy:)}~\approx~~gV\left(\frac{MT}{2\pi}\right)^{3/2}~e^{-M/T},~~~~~~~~
\textrm{\small(light:)}~\approx~~gV\frac{2\pi^2}{4\cdot 45} T^3
$$
In the confined phase the total entropy $s$ produced is proportional to $T^3$. Production of light strange (and antistrange) particles however (with $m_h \gg T$), is proportional to $T^{3/2} \exp\left(-m_h/T\right)$. Thus, the ratio $S$ of strangeness plus antistrangeness production to entropy follows the relation: 
$$ \frac{S}{s}~ \propto ~T^{3/2} \exp\left(-m_K/T\right)$$
 and quickly increases with the collision energy.
In the QGP phase, due to the lower mass  of strange quarks ($m_s<T$), the strangeness yield becomes approximately proportional to the entropy (both $\propto T^3$) and the strangeness to entropy production ratio becomes constant with energy.

Therefore we expect a "jump" in the energy dependence from the quickly increasing strangeness yield for confined matter to the equilibrium value expected for deconfined matter.
The non-monotonic energy dependence of the strangeness to entropy production ratio followed by a plateau at higher energies is expected as a direct consequence of the onset of deconfinement.
Exact calculations within the SMES model \cite{smes} show higher values of the discussed ratio in the hadronic medium at the onset of deconfinement resulting in a very specific shape of the energy dependence, called the ``horn''. (see: \pb data-points in fig. \ref{Fig:horns}).

\section{Results on strangeness production}
\begin{figure}[htb]
	\centering
	\includegraphics[width=0.425\textwidth]{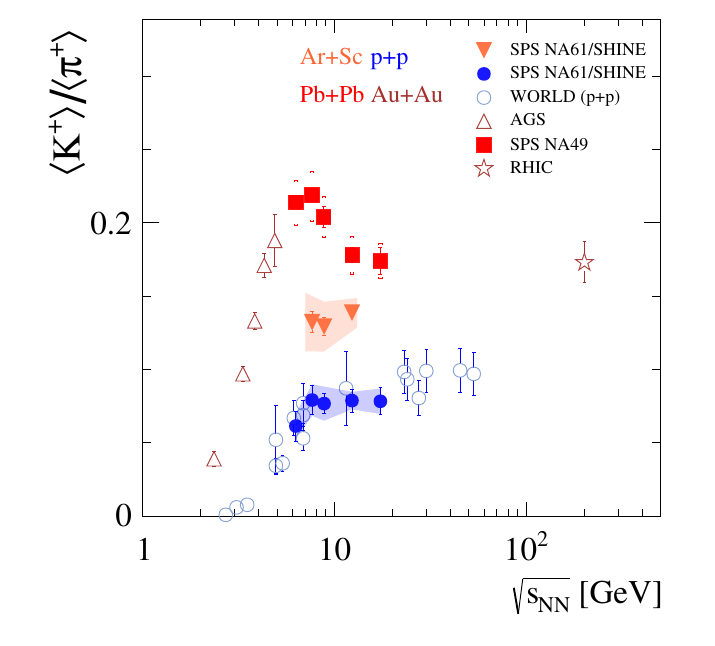}
	\includegraphics[width=0.425\textwidth]{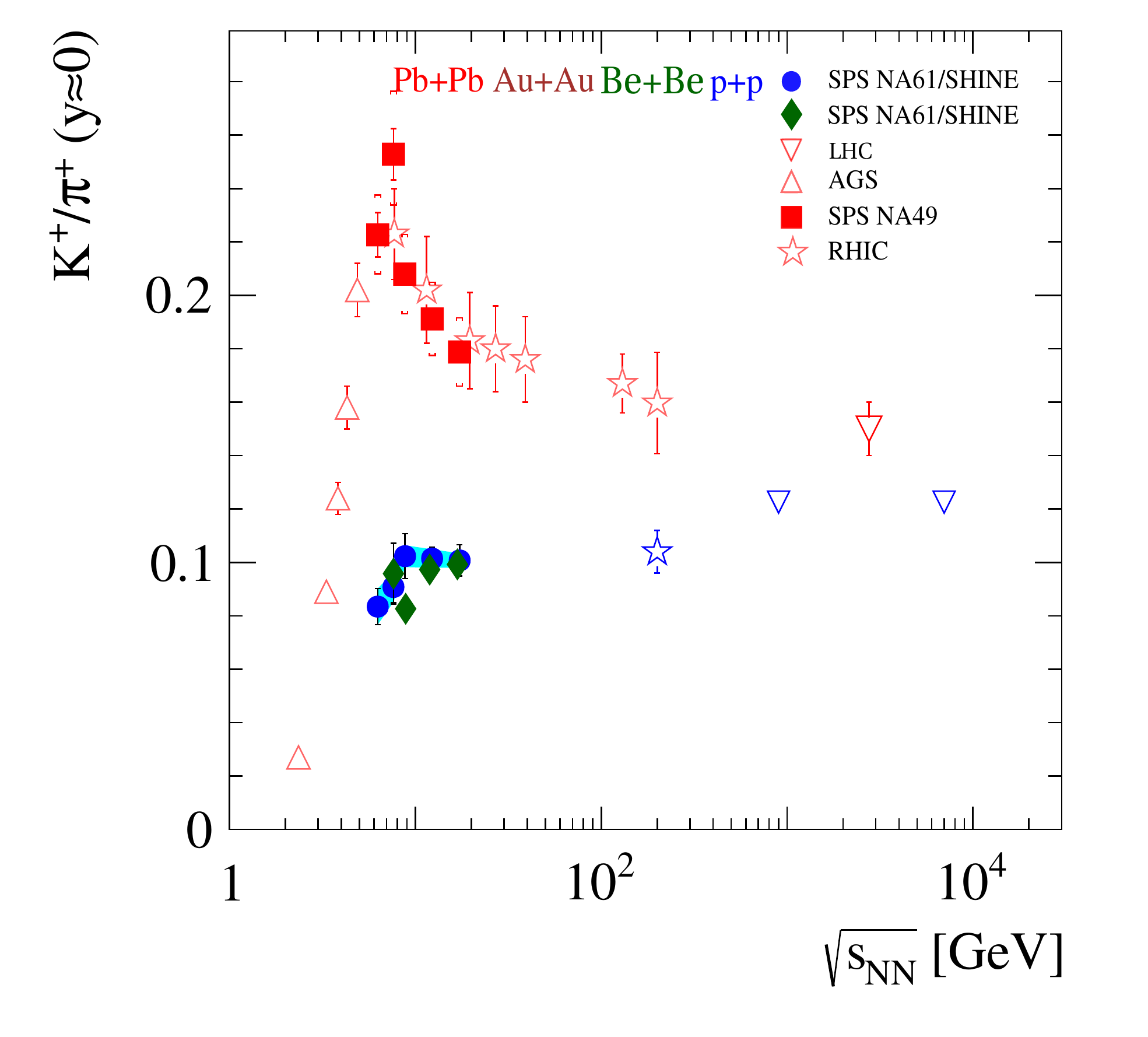}
	\caption{Measurements of $K/\pi$ ratios (at mid-rapidity and in 4$\pi$ acceptance) as a probe of system size dependence of strangeness production. Preliminary results on \ar lie between light and heavy systems. Data on \be is very close to \pp.
	}
	\label{Fig:horns}
\end{figure}

Particle production changes rapidly with collision energy in the vicinity of the onset of deconfinement. One observes a clear, qualitative difference between results obtained in \pp collisions \cite{pp} and collisions of heavy nuclei (\pb or \au) \cite{NA49PbPb, NA49PbPb2, star}. This difference is especially pronounced in the production of strange hadrons as exemplified by the ``horn plot'' shown in Fig. \ref{Fig:horns}.

The NA61/SHINE experiment extends the set of experimental results on hadron production near the onset of deconfinement by unique measurements of collisions of intermediate size nuclei: \be, \ar and Xe+La. The preliminary results released by the collaboration include spectra and multiplicities of identified hadrons in \be \cite{BeBe,antoni} and \ar \cite{naskret2,cpod17} interactions. Additionally, reference measurements of \pp collisions were obtained and are shown as well \cite{antoniPP}.

\begin{figure}[htb]
	\centering
	\includegraphics[width=0.85\textwidth]{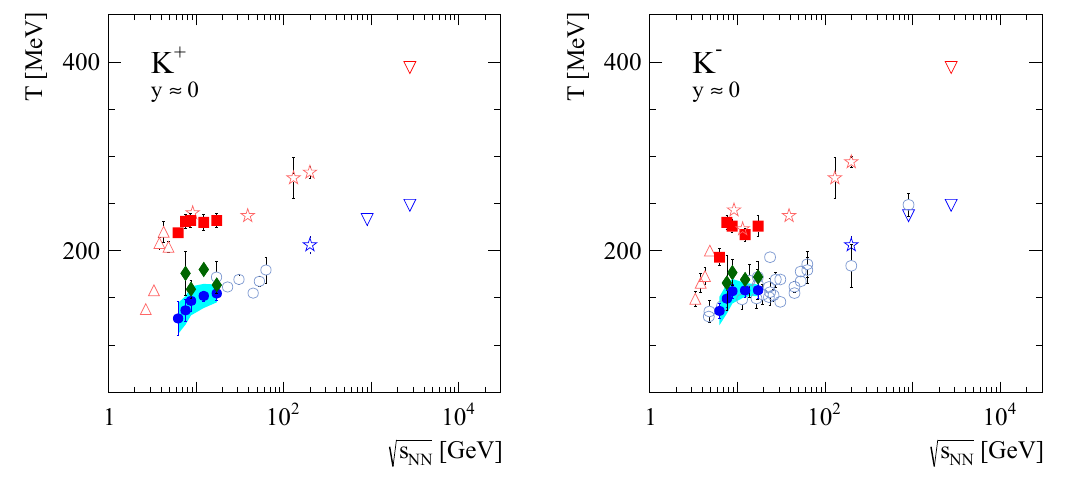}
	\caption{Inverse slope parameter measured in mid-rapidity spectra of positive (left) and negative (right) kaons. The collisions of heavy ions (\pb, \au~-- red points) and \pp (blue points) both show a step-like energy dependence, but at different levels. Recent preliminary measurements of \be fall very close to the clustered points of \pp.
	}
	\label{Fig:T}
\end{figure}

The transverse mass spectra of kaons recorded in \be collisions were fitted with exponential function to extract the inverse slope parameter (at mid-rapidity). \be measurements (green diamonds in fig. \ref{Fig:T}) fall very close to the inverse slope parameter obtained for \pp interactions. Interestingly, a step structure is found like in Pb+Pb collisions, although at a lower level. The measured results for the $\langle K^+ \rangle / \langle \pi^+ \rangle$ ratio at mid-rapidity (Fig.~\ref{Fig:horns}, right panel) in \be and \pp collisions are also very similar -- \be data points are only slightly higher.

The results on the $\langle K^+ \rangle / \langle \pi^+ \rangle$ ratio (in 4$\pi$ acceptance) measured for \ar collisions lie between results on \pp and heavy ions (\pb, \au) interactions -- see Fig.~\ref{Fig:horns}. Data points on \ar are approaching data on \pb towards higher collision energies. It is worth noting, that other measurements of hadron production properties: multiplicity fluctuations \cite{andrey}, pion production \cite{naskret2} and spectra shapes \cite{cpod16,cpod17} show further similarities of \ar and \pb systems.

\section{System size dependence}
From the results described in the previous section one can deduce a clear global trend in data on intermediate size systems. Particle production properties recorded for \be collisions show a close similarity to \pp interactions, while collisions of \ar are much closer to data on \pb.

Such behavior cannot be reproduced so far by any of the models of heavy-ion collisions. Both, statistical and dynamical approaches, fail to describe the new preliminary data provided by NA61/SHINE.

\begin{figure}[htb]
	\centering
	\includegraphics[width=0.35\textwidth, height=0.3\textwidth]{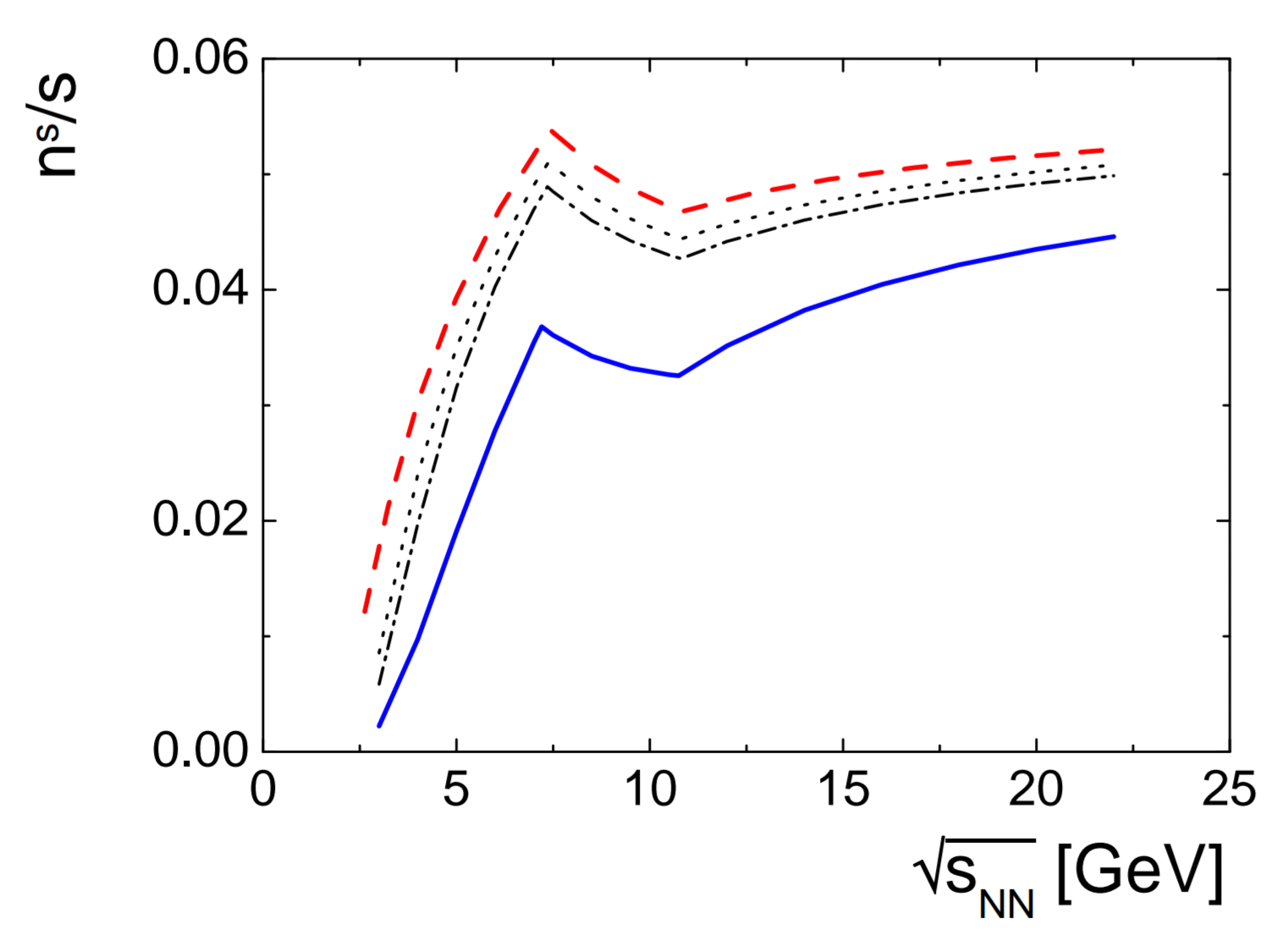}
	\vspace{0.4cm}
	\includegraphics[width=0.3\textwidth, , height=0.3\textwidth, trim={1.cm, 0 15cm 0}, clip]{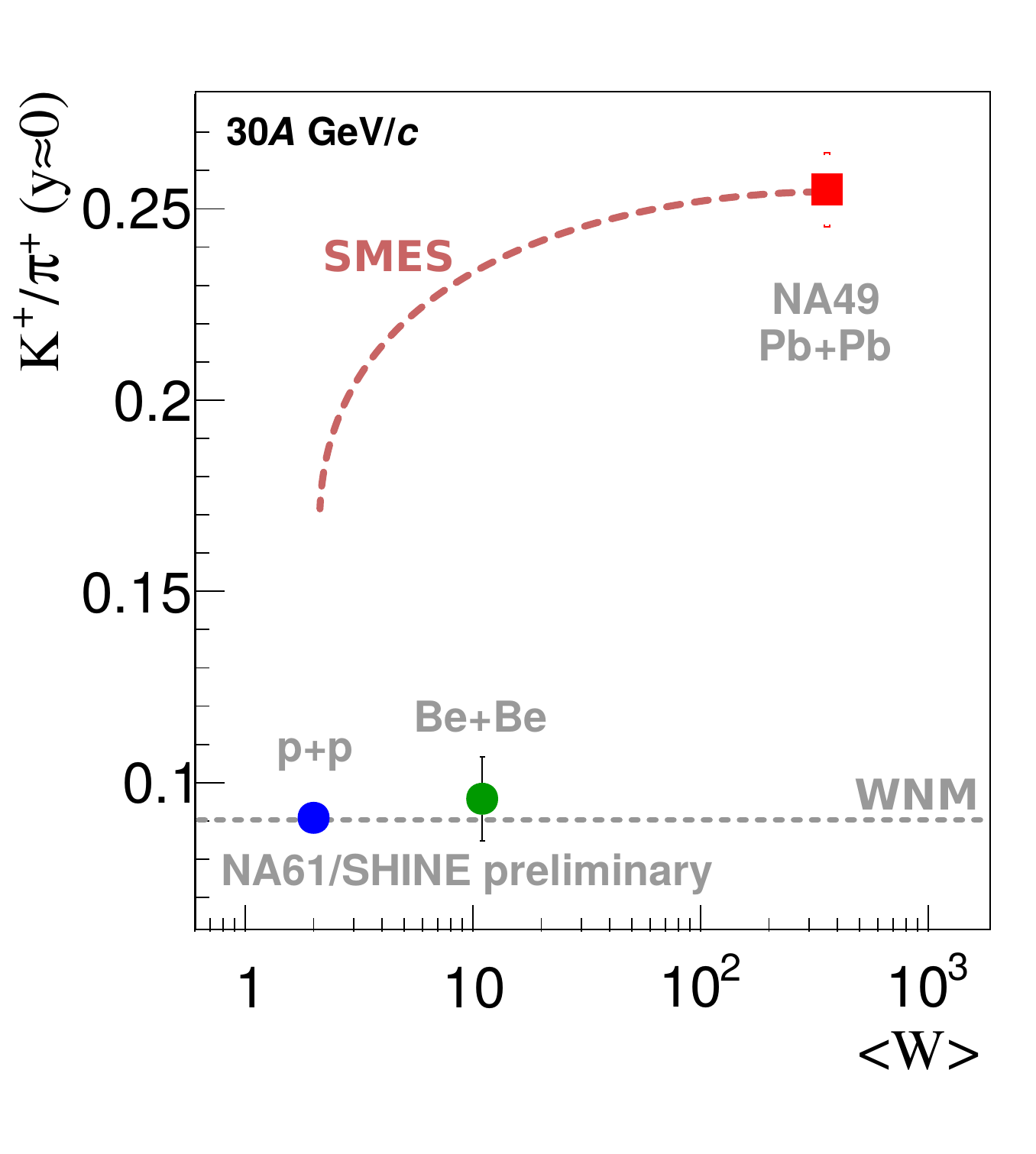}
	\vspace{0.1cm}
	\caption{SMES predicts very different system size dependence of the $K^+/\pi^+$ ratio from the one
		measured by the \na~ experiment.
	}
	\label{Fig:smes_size}
\end{figure}

The system size dependence predicted by SMES is governed by the diminishing effect of the canonical strangeness suppression with increasing volume within statistical models. According to SMES, in case of the \be system, the effect of canonical suppression should already be very small. The contrary is visible in the data -- the results from the \be system are within the statistical errors near the expectations of the wounded nucleon model, which treats each binary nucleon-nucleon collision independently (see Fig.~\ref{Fig:smes_size} for comparison).

\begin{figure}[htb]
	\centering
	\includegraphics[width=0.38\textwidth, height=0.36\textwidth]{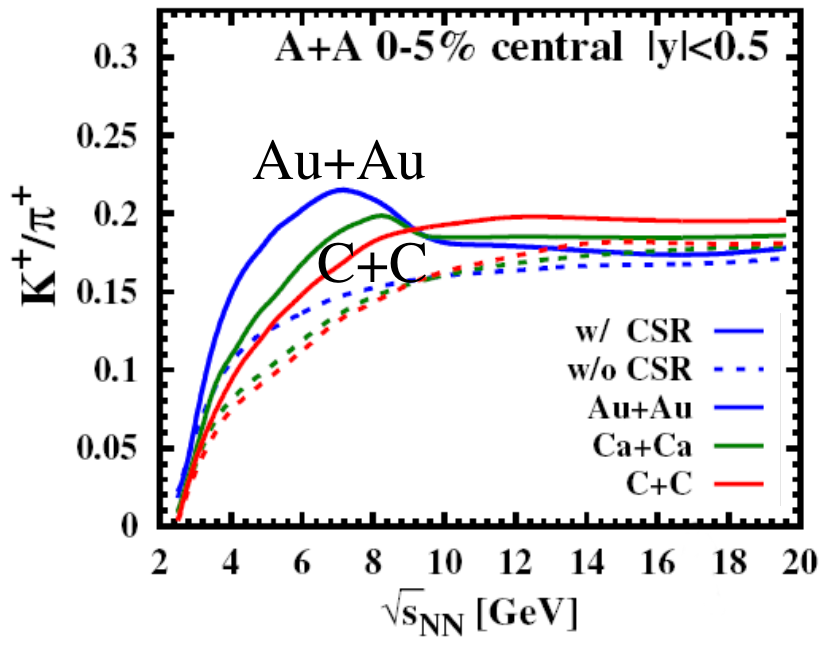}
	\vspace{0.4cm}
	\includegraphics[width=0.38\textwidth, height=0.36\textwidth, trim={15cm, 0 1cm 1.2cm}, clip]{utrecht_3.pdf}
	\caption{PHSD predicts increase of strangeness production with system size at low collision energies ($<$10 GeV) and decrease at high collision energies ($>$10 GeV). PHSD predictions disagree with data at high energies.
	}
	\label{Fig:phsd_size}
	\begin{tikzpicture}[overlay, remember picture]
	\node (1) [] at (-6.25,-0.35) {};
	\draw[line width=0.3ex, gray, opacity=0.9, style=dashed]
	($(1.north)+(5.2, 4.7)$) -- ($(1.north)+(5.2, 5.7)$);
	\node (2) [] at ($(1.north)+(5.0,5.8)$) {\scriptsize \textcolor{gray}{\bfseries 158\textit{A} GeV/\textit{c}}};
	\end{tikzpicture}
\end{figure}

The parton-hadron-string-dynamics model (PHSD) \cite{phsd} is based on a dynamical quasiparticle model (DQPM) matched to reproduce lattice QCD results in thermodynamical equilibrium. The recent update of the model includes a chiral phase transition -- a decreasing mass of strange degrees of freedom already in the confined phase. Only with this update the model is able to qualitatively reproduce the ``horn" in the $K^+/\pi^+$ ratio. However, the PHSD predicts an increase of said ratio with decreasing system size for collision energies larger than $\sqrt{s_{NN}}\approx$ 9 GeV. This is in conflict with the data recently measured by the NA61/SHINE Collaboration (see: Fig.~\ref{Fig:phsd_size}).

\section{Conclusions}
\setlength{\parindent}{2em}
\setlength{\parskip}{0em}
The new set of preliminary results from the NA61/SHINE Collaboration on hadron production in collisions of intermediate size nuclei (\be and \ar) provides a fresh look into the properties of matter in the vicinity of the onset of deconfinement.
The key observation is the clear qualitative difference between the results on small nuclei collisions (\pp, \be) and those of larger sizes (\ar, \pb). In particular, the measurements of transverse spectra and the $K^+/\pi^+$ ratio (at $y$$\approx$0) for \be collisions fall very close to the results obtained for \pp, while the $\langle K^+ \rangle / \langle \pi^+ \rangle$ ratio (in 4$\pi$ acceptance) measured for \ar collisions falls closer to the results from \pb interactions.

\begin{figure}[htb]
	\centering
	\ig[width=0.63\textwidth]{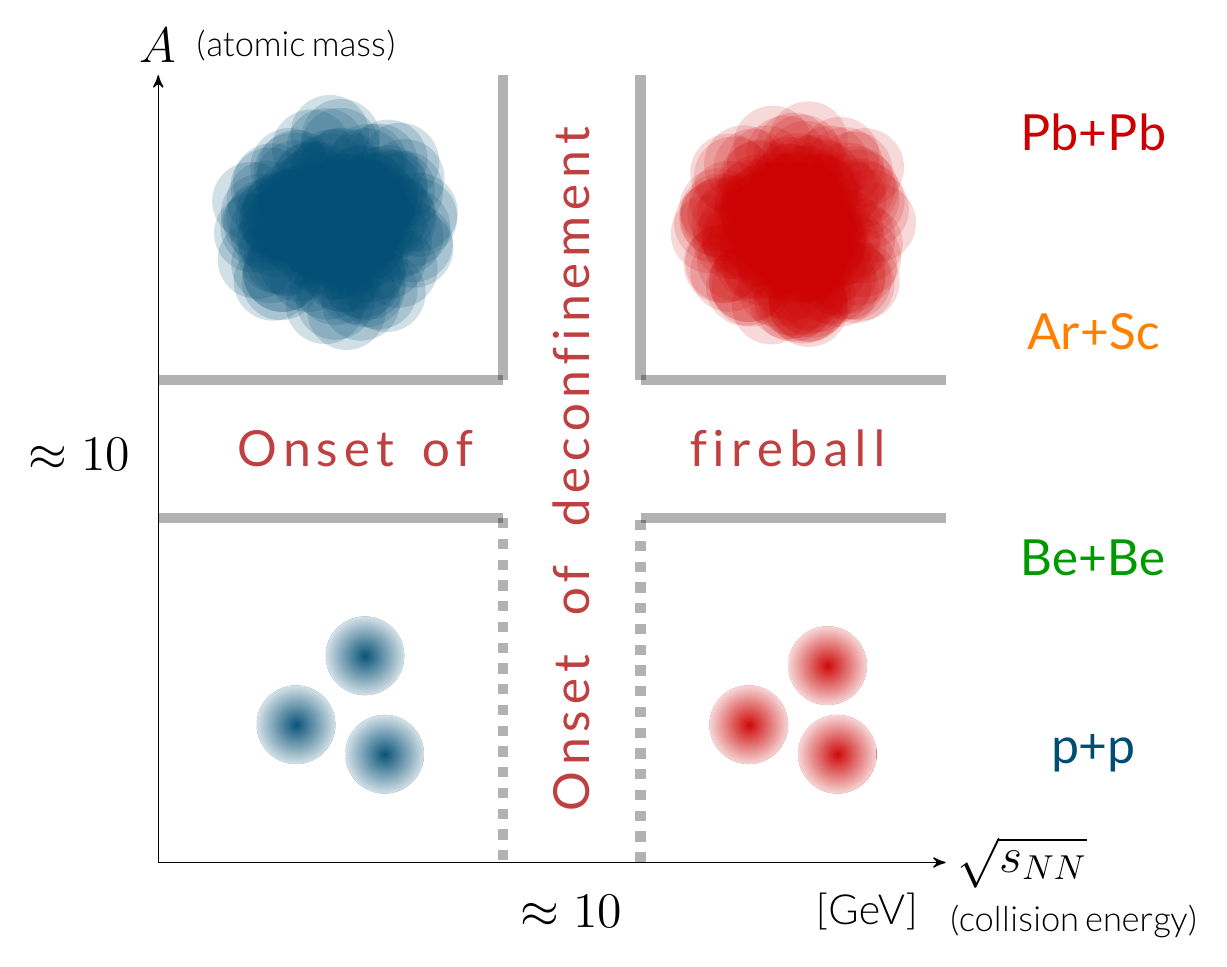}
	\caption{Onset of deconfinement: beginning of creation of QGP with increasing collision energy ($\sqrt{s_{NN}}$). Onset of fireball: beginning of creation of large clusters of strongly interacting mater in A+A collisions with increasing nuclear mass number (A).
	}
	\label{Fig:domains}
\end{figure}

Similar (and even more pronounced) trends are observed in other \linebreak \na measurements, namely: multiplicity fluctuations and $\pi^-$ spectra. This leads to the question whether there might be a microscopic phenomenon leading to such a clear distinction between lighter (\pp, \be) and heavier (\ar, \pb) systems. One may hypothesize that in the first case we seem to observe reactions of independent binary N+N collisions, while in the latter the system evolves collectively as a whole. Figure~\ref{Fig:domains} shows a schematic picture of the two onsets present in heavy-ion collisions: the onset of deconfinement and the the onset of fireball. We can distinguish four hypothetical domains in which hadron production properties should differ significantly.

More experimental data and more numerical tests are needed to determine the precise mechanisms behind the apparent thresholds present in the system size dependence of particle production. No theoretical models can as yet reproduce the experimental results, which clearly puts a challenge on the market, especially considering more data on intermediate size systems coming from the NA61/SHINE Collaboration in the near future.

\end{document}